\documentclass[a4paper,11pt]{article}
\usepackage{pos}

\title{Study of the long-term Very High Energy emission of the blazars 1ES 1215+3031 and VER J0521+211 with the HAWC gamma-ray observatory}
 \ShortTitle{Study of the  VHE emission of 1ES 1215+3031 and VER J0521+211 with HAWC}

\author*[a]{Fernando Ureña-Mena}
\author[a]{Alberto Carramiñana}

\author[b]{Anna Lia Longinotti}
\author[a]{Daniel Rosa-González}

\affiliation[a]{Instituto Nacional de Astrofísica, Óptica y Electrónica (INAOE)\\
  Luis Enrique Erro 1, Tonantzintla, Puebla, Mexico}

\affiliation[b]{Instituto de Astronomía, Universidad Nacional Autónoma de México (IA-UNAM),\\ Ciudad de México, Mexico\\
}

\onbehalf{for the HAWC collaboration} 

\emailAdd{furena@inaoep.mx}

\abstract{Abstract: Blazars are the most abundant type of extragalactic gamma-ray source, usually presenting high variability across the electromagnetic spectrum. Their Very High Energy (VHE, above 0.1 TeV) emission has been studied in detail using Air Cherenkov Telescopes, with observations biased to flaring periods while their average activity has not been properly characterized. In this work, we report the results of 2090 days of quasi-continuous observations of the blazars 1ES 1215+303 and VER J0521+211, carried out with the High Altitude Water Cherenkov (HAWC) gamma-ray observatory. Fitting a power-law attenuated by photon-photon interaction with the extragalactic background light, we obtained a 6.2 $\sigma$ level detection for 1ES 1215+303 and a 4.3 $\sigma$ marginal detection for VER J0521+211. With the inclusion of the HAWC TeV spectrum, we built quasi-simultaneous multiwavelength spectral distributions and fitted a leptonic emission model to the observed data. }

\ConferenceLogo{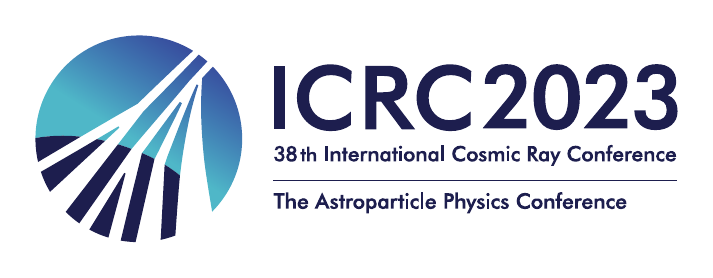}

\FullConference{%
38th International Cosmic Ray Conference (ICRC2023)\\
  26 July - 3 August, 2023\\
  Nagoya, Japan}

\begin{document}
\maketitle

\section{Introduction}

Blazars are a type of radio loud Active Galactic Nuclei (AGN), whose relativistic jets point nearly directly to the observer's line of sight. A subclass of blazars named BL Lac objects constitute the most abundant extragalactic  sources in the gamma ray sky \cite{tevcat2008}. Their TeV emission is due  to the relativistic beaming in the jet, that increases their apparent luminosity \cite{dermer2009}. As for any other extragalactic TeV source, their emission is attenuated by photon-photon interaction with the extragalactic background light (EBL)\cite{dwek2013,albert2022}.\\

The Spectral Energy Distributions (SED) observed in BL Lac objects usually presents a two-peaked structure \cite{fossati1998}, with a low energy component from radio to X-rays and a high energy component from X-rays to gamma rays. The low-energy component is generally modeled as produced by synchrotron emission from an accelerated electron population in the relativistic jet, which contains a strong magnetic field. On the other hand,  the most popular explanation for the second component is  Compton  up-scattering of synchrotron photons by the same electron population, which is called Synchrotron Self Compton (SSC)  scenario \cite{finke2008}.\\

The High Altitude Water Cherenkov (HAWC) observatory is a gamma-ray and cosmic ray detector located at 4100 m above sea level  in Puebla, Mexico. Due to its duty cycle, HAWC provides almost continuous observations allowing the characterization of long-term  TeV gamma-ray emission. A search for gamma ray emission from AGNs in 1523 days of HAWC data was recently published \cite{albert2021}, reporting two BL Lac objects with a solid detection (Mrk 421 and Mrk 501) and two with a significance between $3 \sigma$ and $ 5 \sigma$, 1ES 1215+303 and VER J0521+211.\\ 

The blazar 1ES 1215+303 (R.A.$=12^{h}17^{m}48.5^{s}$, Dec.$=	+30^{\circ}06^{'}06^{''}$) is a well established MeV-GeV and TeV source \cite{alecksic2012,valverde2020}, with a measured redshift of $z=0.130$ \cite{paiano2017}. It is located near the Coma Cluster region, with an angular separation of just $0.88$ from another gamma ray blazar, PG 1218+304 (R.A.$=12^{h}21^{m}26.3^{s}$, Dec.$=	+30^{\circ}11^{'}29^{''}$,$Z=0.182$). In the HAWC survey \cite{albert2021}, this source is reported with a significance of $3.6\sigma$. \\

The blazar VER J0521+211 (R.A.$=05^{h}21^{m}45^{s}$, Dec.$=	+21 ^{\circ}12^{'}51.4^{''}$, $z=0.108$ \cite{shaw2013}) is also a well known gamma ray source \cite{archambault2013,adams2022}. It is located close to the galactic plane,  just 3.08 degrees away from the Crab Nebula. In this case, the significance reported for the 1523-day HAWC data set was $3.2 \sigma$ \cite{albert2021}.

\section{Data}

We have recently analyzed 2090 days of HAWC data for the blazars 1ES 1215+303 and VER J0521+211, which due to their convenient declinations $|{\delta - 19 ^{\circ}}|\lesssim 10^{\circ}$ present long daily transits ($>6$ hours). The data set observations covered from 2014 November 26 to 2021 January 14. \\

We also built  multiwavelenght SEDs from radio to gamma rays. For this purpose, we used simultaneous Fermi-LAT data in the energy range (100 MeV-500 GeV), as well as archival quasi-simultaneous observations for the rest of the SED \cite{valverde2020,acciari2020,lister2018,sotnikova2022} and our HAWC results. Finally,  we fit a physical model to the SED in order to determine if it is consistent with the VHE emission. 

\section{Methodology}

We fit the following model to the observed HAWC spectra:
\begin{equation}
\label{eq:spc}
\left(\frac{\mathrm{d}N}{\mathrm{d}E}\right)_{obs}=K \left(\frac{E}{1 \ \mathrm{TeV}}\right)^{-\alpha}e^{-\tau(E,z)},
\end{equation}
 where $K,\alpha$ are our two fitting parameters. The term $e^{-\tau(E,z)}$ accounts for the EBL attenuation and follows the model presented in \cite{dominguez2011}. We used $z=0.13$ for 1ES 1215+303 \cite{paiano2017} and $z=0.108$ for VER J0521+211 \cite{shaw2013}. Then, we computed the test statistic, which is defined as $TS=2\ln{(\mathcal{L}_1/\mathcal{L}_0)}$, where $\mathcal{L}_1,\mathcal{L}_0$ are the likelihood for the best fit point source and null hypothesis, respectively. \\

 We also fit a physical model to the observed SEDs, which included the 2090 day HAWC data. Following some similar works \cite{valverde2020,acciari2020}, we fit a two zone SSC emission model to the data. We assumed the existence of a inner compact region in the jet that we called ``blob" and an outer extended region that we called ``core" for which $R_{core}>R_{blob}$.  The electron population in the ``blob"  is assumed to follow a broken power law spectral distribution: 

 \begin{equation}
 N_e(\gamma^\prime)\propto \begin{cases} \gamma^{\prime^{-p_1}} \text{for } \gamma^\prime<\gamma_c^\prime \\  
     \gamma^{\prime^{-p_2}} \text{for } \gamma^\prime>\gamma_c^\prime
    
    \end{cases},
\end{equation}
where $\gamma$ is the comoving electron Lorentz factor, $p_1,p_2$ are the broken power law indices and $\gamma_c^\prime$ is the break Lorentz factor. In case of the ``core", their spectral electron distribution is assumed to follow a single power law: 

\begin{equation}
N_e(\gamma_e^\prime) \propto \gamma_e^{\prime-p},
\end{equation}

We assume that these electron populations are immerse in a randomly oriented magnetic field with mean intensity $B$ and that the emission regions are moving with  Doppler factor $ \delta$, which is defined as:

\begin{equation}
    \delta={\left[\Gamma\left(1-\frac{v}{c}\cos\theta\right)\right]^{-1}},
\end{equation}

where $c$ is the speed of light, $\Gamma$ is  the Lorentz factor, $v$ is velocity of the jet and $\theta$ is the viewing angle of the region with respect to the observer's line of sight. 

We simulated the SSC emission using the software package \texttt{agnpy} \cite{nigro2022} and obtained the best fit value for the fitting parameters using the software package \texttt{iminuit} \cite{dembinski2020}. The parameters $R_{core}$  and $R_{blob}$ has been fixed to values obtained from previous works \cite{valverde2020,acciari2020}, which are shown in Table \ref{tab:best_fit_values}.

\section{Results and Discussion}

The values that we obtained for the gamma ray spectral parameters defined in Eq \ref{eq:spc} are shown in Table \ref{tab:table1}. We observe an increase in TS with respect to 1523 day data for both sources, consistent with a solid detection for 1ES 1215+303 ($\sqrt{TS}>5$) and more significant marginal detection for VER J0521+211. However, these results are still preliminary because we have not considered the likely contamination from PG 1218+304, which lies at just 0.88 degrees from 1ES 1215+303, and from the Crab Nebula, at just 3.07 degrees from VER J0521+211. We will include  that part of the analysis of the analysis in our final publication. \\

\begin{table}
\centering
\caption{Gamma-ray spectral parameters}
\label{tab:table1}
\begin{tabular}{@{}lccc@{}}\hline
\textbf{Source} & {$K$ ($\times10^{-12}$ TeV$^{-1}$cm$^{-2}$s$^{-1}$)} & \textbf{$\alpha$} & $TS$ \\ \hline
1ES 1215+303 & $ 2.09\pm 0.69$ & $3.46 \pm 0.16$ & $43.9 \ (6.6\sigma)$ \\
VER J0521+211 & $2.27 \pm 0.56 $ & $2.89 \pm 0.21 $ & $18.3 \ (4.3\sigma)$\\ \hline
\end{tabular}
\end{table}

\begin{figure}
	\centering
    \includegraphics[width=\textwidth]{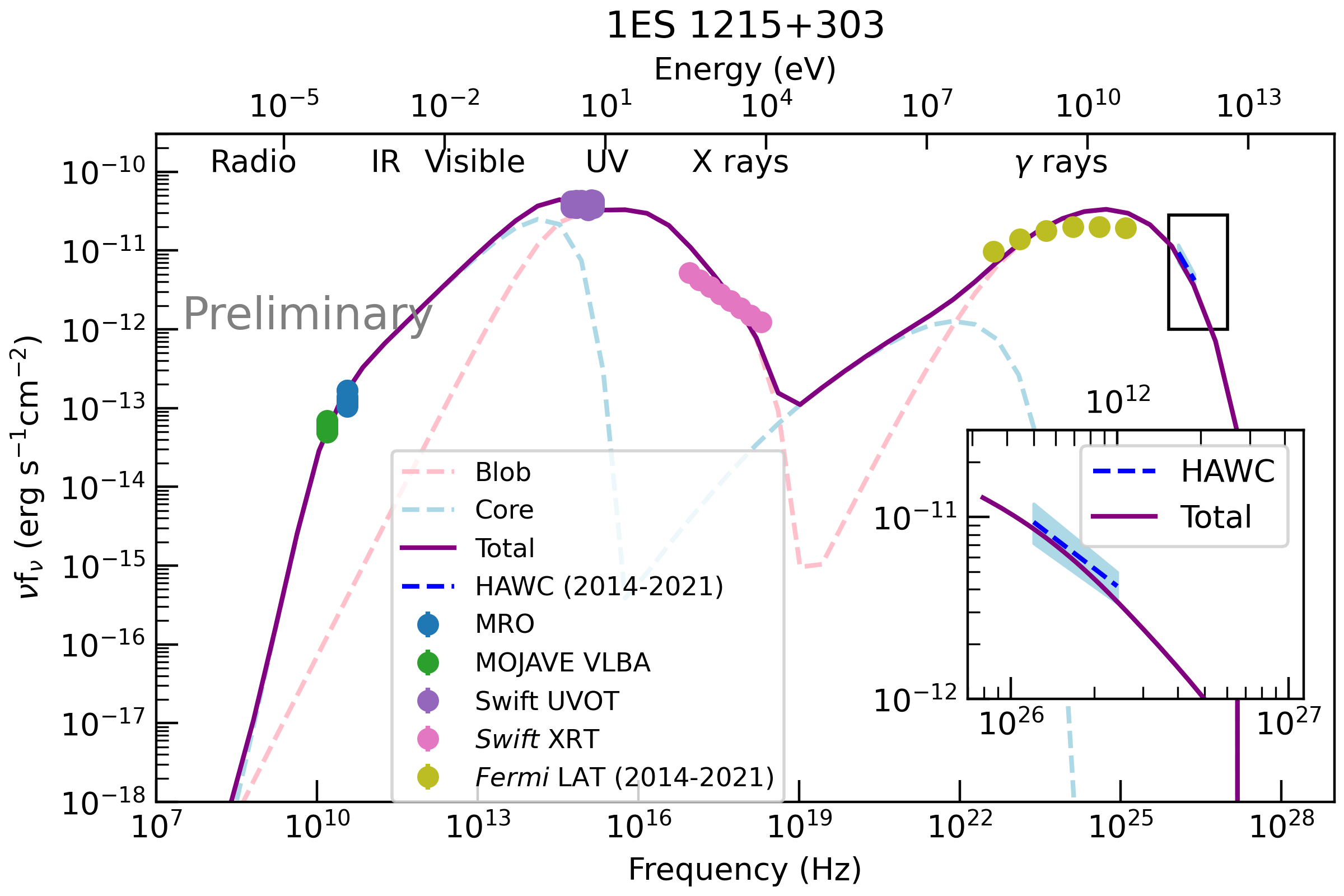}
    \caption{Multiwavelength spectral energy distribution of the blazar 1ES 1215+303 with leptonic model fit. The dashed pink curve depicts the emission from inner zone (blob) and the dashed blue curve shows the emission from the outer zone (core). The light blue region corresponds to the HAWC 1$\sigma$ error band.}
    \label{fig:1ES}
\end{figure}

\begin{figure}
	\centering
    \includegraphics[width=\textwidth]{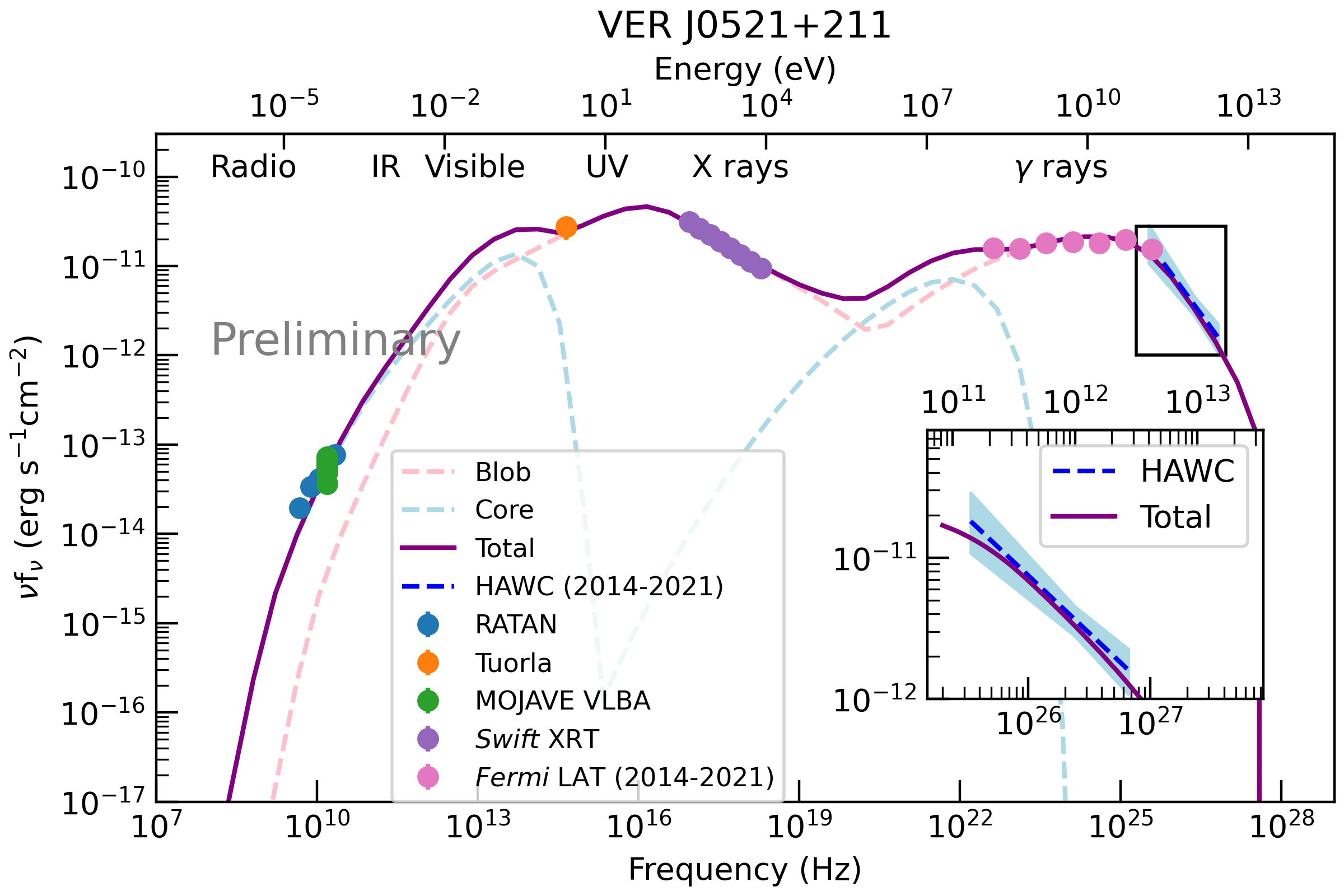}
    \caption{Multiwavelength spectral energy distribution of the blazar VER J0521+211 with two-zone leptonic model fit. The dashed pink curve depicts the emission from inner zone (blob) and the dashed blue curve shows the emission from the outer zone (core).  The light blue region corresponds to the HAWC 1$\sigma$ error band.}
    \label{fig:VER}
\end{figure}

The results for some  best fit values of the physical model are shown in Table \ref{tab:best_fit_values}, which are consistent with what is expected for this kind of sources (see for example \cite{acciari2020}). Figures \ref{fig:1ES} and \ref{fig:VER}  show the quasi-simultaneous SEDs together with the best fit two-zone model that confirm that this scenario is able to explain the long-term VHE emission observed by HAWC.

\begin{table}
\centering
\caption{Best fit values for some model parameters}
\label{tab:best_fit_values}
\begin{tabular}{@{}lcc@{}}
\hline
\textbf{Parameter} &  VER J0521+211 & 1ES 1215+303 \\ \hline
 Blob radius (fixed) $R_b^\prime$ (cm)& $5.1\times10^{16}$
  & $1.35\times10^{16}$ \\
Magnetic Field (blob)$B$ (G) &  $0.110\pm0.009$ & $0.021\pm0.001$ \\ 
Doppler Factor (blob) $\delta$ & $30.0\pm0.9$ & $30.0\pm0.9$ \\
Core radius (fixed) $R_b^\prime$ cm& $1\times10^{17}$
 & $3.7\times10^{17}$\\
Magnetic Field (core)$B$ (G) &  $0.011\pm0.001$& $0.035\pm0.003 $ \\ 
Doppler Factor (core) $\delta$  &  $ 11.8\pm0.4$ & $30.0\pm1.1 $ \\

 \hline
\end{tabular}
\end{table}

\section{Conclusions}

Using 2090 days of HAWC data,we observe a promising increase in the significance of two HAWC blazars with respect to the result from the 1523 day data set. Our results are consistent with a 6.2 $\sigma$ detection for 1ES 1215+303 and a $4.3\sigma$ marginal detection for VER J0521+211. However, these observations are likely contaminated from other well known gamma-ray sources, whose effect will be considered in our final publication. \\

We fit a two-zone leptonic model to quasi-simultaneous SEDs of these two objects. We confirm that this scenario can explain the long-term VHE emission as characterized by HAWC.

\clearpage
\section*{Full Authors List: HAWC Collaboration}
%
%\noindent \textbf{Note comment afterwards:} Collaborations have the possibility to provide an authors list in xml format which will be used while generating the DOI entries making the full authors list searchable in databases like Inspire HEP. \\
%
\scriptsize
\noindent
%first.author$^1$, 
%second.author$^2$, 
%third.author$^3$ % .... more names
%and 
%last.author$^{n}$ \\
%
%\noindent
%$^1$first.affiliation.
%$^2$second.affiliation. % .... more affiliation
%$^{m}$last.affiliation.
\vskip2cm
\noindent
A. Albert$^{1}$,
R. Alfaro$^{2}$,
C. Alvarez$^{3}$,
A. Andrés$^{4}$,
J.C. Arteaga-Velázquez$^{5}$,
D. Avila Rojas$^{2}$,
H.A. Ayala Solares$^{6}$,
R. Babu$^{7}$,
E. Belmont-Moreno$^{2}$,
K.S. Caballero-Mora$^{3}$,
T. Capistrán$^{4}$,
S. Yun-Cárcamo$^{8}$,
A. Carramiñana$^{9}$,
F. Carreón$^{4}$,
U. Cotti$^{5}$,
J. Cotzomi$^{26}$,
S. Coutiño de León$^{10}$,
E. De la Fuente$^{11}$,
D. Depaoli$^{12}$,
C. de León$^{5}$,
R. Diaz Hernandez$^{9}$,
J.C. Díaz-Vélez$^{11}$,
B.L. Dingus$^{1}$,
M. Durocher$^{1}$,
M.A. DuVernois$^{10}$,
K. Engel$^{8}$,
C. Espinoza$^{2}$,
K.L. Fan$^{8}$,
K. Fang$^{10}$,
N.I. Fraija$^{4}$,
J.A. García-González$^{13}$,
F. Garfias$^{4}$,
H. Goksu$^{12}$,
M.M. González$^{4}$,
J.A. Goodman$^{8}$,
S. Groetsch$^{7}$,
J.P. Harding$^{1}$,
S. Hernandez$^{2}$,
I. Herzog$^{14}$,
J. Hinton$^{12}$,
D. Huang$^{7}$,
F. Hueyotl-Zahuantitla$^{3}$,
P. Hüntemeyer$^{7}$,
A. Iriarte$^{4}$,
V. Joshi$^{28}$,
S. Kaufmann$^{15}$,
D. Kieda$^{16}$,
A. Lara$^{17}$,
J. Lee$^{18}$,
W.H. Lee$^{4}$,
H. León Vargas$^{2}$,
J. Linnemann$^{14}$,
A.L. Longinotti$^{4}$,
G. Luis-Raya$^{15}$,
K. Malone$^{19}$,
J. Martínez-Castro$^{20}$,
J.A.J. Matthews$^{21}$,
P. Miranda-Romagnoli$^{22}$,
J. Montes$^{4}$,
J.A. Morales-Soto$^{5}$,
M. Mostafá$^{6}$,
L. Nellen$^{23}$,
M.U. Nisa$^{14}$,
R. Noriega-Papaqui$^{22}$,
L. Olivera-Nieto$^{12}$,
N. Omodei$^{24}$,
Y. Pérez Araujo$^{4}$,
E.G. Pérez-Pérez$^{15}$,
A. Pratts$^{2}$,
C.D. Rho$^{25}$,
D. Rosa-Gonzalez$^{9}$,
E. Ruiz-Velasco$^{12}$,
H. Salazar$^{26}$,
D. Salazar-Gallegos$^{14}$,
A. Sandoval$^{2}$,
M. Schneider$^{8}$,
G. Schwefer$^{12}$,
J. Serna-Franco$^{2}$,
A.J. Smith$^{8}$,
Y. Son$^{18}$,
R.W. Springer$^{16}$,
O.~Tibolla$^{15}$,
K. Tollefson$^{14}$,
I. Torres$^{9}$,
R. Torres-Escobedo$^{27}$,
R. Turner$^{7}$,
F. Ureña-Mena$^{9}$,
E. Varela$^{26}$,
L. Villaseñor$^{26}$,
X. Wang$^{7}$,
I.J. Watson$^{18}$,
F. Werner$^{12}$,
K.~Whitaker$^{6}$,
E. Willox$^{8}$,
H. Wu$^{10}$,
H. Zhou$^{27}$

\vskip2cm
\noindent
$^{1}$Physics Division, Los Alamos National Laboratory, Los Alamos, NM, USA,
$^{2}$Instituto de Física, Universidad Nacional Autónoma de México, Ciudad de México, México,
$^{3}$Universidad Autónoma de Chiapas, Tuxtla Gutiérrez, Chiapas, México,
$^{4}$Instituto de Astronomía, Universidad Nacional Autónoma de México, Ciudad de México, México,
$^{5}$Instituto de Física y Matemáticas, Universidad Michoacana de San Nicolás de Hidalgo, Morelia, Michoacán, México,
$^{6}$Department of Physics, Pennsylvania State University, University Park, PA, USA,
$^{7}$Department of Physics, Michigan Technological University, Houghton, MI, USA,
$^{8}$Department of Physics, University of Maryland, College Park, MD, USA,
$^{9}$Instituto Nacional de Astrofísica, Óptica y Electrónica, Tonantzintla, Puebla, México,
$^{10}$Department of Physics, University of Wisconsin-Madison, Madison, WI, USA,
$^{11}$CUCEI, CUCEA, Universidad de Guadalajara, Guadalajara, Jalisco, México,
$^{12}$Max-Planck Institute for Nuclear Physics, Heidelberg, Germany,
$^{13}$Tecnologico de Monterrey, Escuela de Ingeniería y Ciencias, Ave. Eugenio Garza Sada 2501, Monterrey, N.L., 64849, México,
$^{14}$Department of Physics and Astronomy, Michigan State University, East Lansing, MI, USA,
$^{15}$Universidad Politécnica de Pachuca, Pachuca, Hgo, México,
$^{16}$Department of Physics and Astronomy, University of Utah, Salt Lake City, UT, USA,
$^{17}$Instituto de Geofísica, Universidad Nacional Autónoma de México, Ciudad de México, México,
$^{18}$University of Seoul, Seoul, Rep. of Korea,
$^{19}$Space Science and Applications Group, Los Alamos National Laboratory, Los Alamos, NM USA
$^{20}$Centro de Investigación en Computación, Instituto Politécnico Nacional, Ciudad de México, México,
$^{21}$Department of Physics and Astronomy, University of New Mexico, Albuquerque, NM, USA,
$^{22}$Universidad Autónoma del Estado de Hidalgo, Pachuca, Hgo., México,
$^{23}$Instituto de Ciencias Nucleares, Universidad Nacional Autónoma de México, Ciudad de México, México,
$^{24}$Stanford University, Stanford, CA, USA,
$^{25}$Department of Physics, Sungkyunkwan University, Suwon, South Korea,
$^{26}$Facultad de Ciencias Físico Matemáticas, Benemérita Universidad Autónoma de Puebla, Puebla, México,
$^{27}$Tsung-Dao Lee Institute and School of Physics and Astronomy, Shanghai Jiao Tong University, Shanghai, China,
$^{28}$Erlangen Centre for Astroparticle Physics, Friedrich Alexander Universität, Erlangen, BY, Germany

%==============

%%%%%%%%%%%%%%%%%%%%%%%%%%%%

\end{document}